\pdfoutput=1

\documentclass[aip,preprint,jcp]{revtex4-1}

\usepackage{graphicx}% Include figure files
\usepackage{braket}
\usepackage{color}

\begin{document}

% \preprint{AIP/123-QED}

\title[]{The $\rm{H_{2}^{+} + He}$ proton transfer reaction: 
  quantum reactive differential cross sections linked with velocity mappings }% Force line breaks with \\
% \\with Forced Linebreak\footnote{Error!}
%\thanks{Footnote to title of article.}

\author{Mario Hern{\'a}ndez Vera}
\author{R. Wester}%
\author{F. A. Gianturco}%
\email{francesco.gianturco@uibk.ac.at.}

\affiliation{ Institut f$\ddot{u}$r Ionenphysik und Angewandte Physik, Universit$\ddot{a}$t Innsbruck,
 Technikerstr.25/3,A-6020,Innsbruck, Austria.}%

\date{\today}% It is always \today, today,
             %  but any date may be explicitly specified

\begin{abstract}
 We construct the velocity map images of the proton transfer reaction 
 between helium and molecular hydrogen ions $\rm{H_{2}^{+}}$. 
 We perform simulations of imaging experiments at one representative total collision energy taking into account 
 the inherent aberrations of the velocity mapping in order to explore the feasibility of direct comparisons 
 between theory and future experiments planned in our laboratory. 
 The asymptotic angular distributions of the fragments in a 3D velocity space is determined from the quantum  
 state-to-state differential reactive cross sections and reaction probabilities which are computed by using 
 the time-independent coupled channel hyperspherical coordinate method. 
 The calculations employ an earlier ab initio potential energy surface computed 
 at the FCI/cc-pVQZ level of theory. The present simulations indicate that the planned experiments 
 would be selective enough to differentiate between product distributions resulting from different 
 initial internal states of the reactants.
%
%Valid PACS numbers may be entered using the \verb+\pacs{#1}+ command.
\end{abstract}

% \pacs{Valid PACS appear here}

\maketitle

\section{\label{sec:intro} Introduction}

  Molecular hydrogen ions (MHIs), in the form of $\rm{H_2}^{+}$ and $\rm{HD}^{+}$, are considered 
  the simplest molecules in nature. This special characteristic makes them important 
  benchmark systems to test different theories describing their fundamental properties.   
  As members of the family of one-electron molecules, they have
  enormous potential in ultraprecise spectroscopy \cite{Korobov:14,Schiller:17}, and consequently, they have been 
  proposed as a means to determine the value of fundamental physical constants, to test relativistic 
  quantum mechanics and QED \cite{Biesheuvel:16}, or in order to perform the construction of  precise molecular 
  clocks~\cite{Schiller:14}.
  In astrophysics, elemental reactions involving MHIs, like the radiative association 
  between $\rm{H^{+}}$ and $\rm{H}$ and the proton exchange between $\rm{HeH^{+}}$ and $\rm{H}$, 
  are key processes to understand the synthesis of the molecular hydrogen in the early universe~\cite{Wolf:13}. 
  They also provide the essential chemical ingredients for assessing the destruction/survival paths for the $\rm{HeH^{+}}$ 
  molecule in the Interstellar Medium \cite{Bovino:11}.
  From the theoretical point of view, investigation of proton exchange reactions have employed frequently
  $\rm{H_{2}}$ or $\rm{H_{2}^+}$  as models to simplify the quantum calculations. 
  Proton exchange reactions between $\rm{H_2}^{+}$/$\rm{HD}^{+}$ and $\rm{H}$ or $\rm{He}$ 
  are part of the reduced group of reactions that can be studied using 
  very accurate Coupled Channel (CC) scattering calculations.

  One important reaction that has been investigated extensively during the last few decades
  is the proton transfer reaction
  between $\rm{H_{2}^{+}}$ and $\rm{He}$~\cite{Fazio:12,Sujitha:12,Sanjay:98,Mahapatra:97,John:90,Tomi:87}.   
  The former reaction is expected to occur adiabatically up to total energies of about 10 eV, 
  making it a prototype case for comparing theory and experiment. 
  Recently, De Fazio et al. \cite{Fazio:12} 
  computed total integral reactive cross section from CC time-independent calculations 
  obtaining good agreement with previous experimental data.  
  Different experiments~\cite{Baer:86,Pijkeren:84,Pollard:91,Cohen:91,Tang:02} have demonstrated that 
  the energy stored in the vibrational modes of 
  the \emph{ortho}-hydrogen (o-$\rm{H_{2}^+}$) or \emph{para}-hydrogen (p-$\rm{H_{2}^+}$) 
  molecular ions  are more efficient in promoting the
  reaction than the translational energy. 
  This behaviour has been reproduced by CC calculations ~\cite{Fazio:12} and even by the simpler 
  infinite-order sudden approximation (RIOSA) treatments \cite{Baer:86}.    
  Besides this, we have found no specific calculations using accurate CC 
  reactive approaches that have analysed 
  the angular distributions from the corresponding reactive differential cross sections (DCSs). 
  It is the aim of the present paper to examine in detail just such specific dynamical observables, 
  as we shall discuss below.

  Velocity map imaging experiments have proved to be very effective for studying 
  the elementary mechanisms which act 
  in gas-phase ion-molecule reactions. The title reaction is indeed among the most suitable candidates for this 
  kind of experimental analysis owing to the clear mass differences between the charged fragments. 
  The additional and significant advantage of this technique is its capacity of resolving individual 
  reaction products irrespective of their scattering angle or velocity magnitude 
  \cite{Wester:14,Carrascosa:2017,Eduardo:15}. It is also possible to employ this method to determine 
  scattering branching ratios which are indeed critical observables for the case of proton exchange 
  reactions involving the $\rm{HD}^{+}$ ion. One should also note here that during the last decade the 
  experimentally achievable intensity of rates of product ions have increased by orders of magnitude, and 
  therefore much more detailed investigations, with better statistical and systematic accuracy 
  for the collected data, can now be performed~\cite{Meyer:2017}.  However, usually the resolution 
  is not yet sufficient enough to distinguish among the rotational energy spacings appearing between the 
  3D Newton spheres formed by the products, and the resulting mixture in the velocity space needs to be 
  very carefully unraveled  in the ensuing  spectra analysis.  
  Therefore, significant numerical simulations of the experiment velocity maps should also realistically include 
  the existing resolution of the crossed-beam imaging data that is 
  determined by  the aberration present in  the velocity mapping and the actual angular 
  spread of the two reacting beams. This aspect of the experimental situation will therefore be explicitly included 
  in  our present numerical simulations in order to better link our computational findings with possible experimental observations.

  With regard to the object of the present study, i.e. the proton transfer reaction 
  between $\rm{H_2}^{+}$ and $\rm{He}$, 
  \begin{equation} \label{eq:reaction}
  \rm{He + H_{2}^{+}(v,j) \to HeH^{+}(j',v') + H} , 
  \end{equation} 
  we would like to answer in the main, and among others, the following specific question: 
  Is it possible to distinguish  among experimental images of reactions associated with different  
  initial rotational quantum states of the reactants?  
  The collision energy and the initial quantum state of the reactant determine the final 
  angular distributions of the products in the velocity map (VM). 
  Therefore, realistic simulations that would include the intrinsic dispersion of the imaging 
  spectrometer are useful to assess more directly the influence on the final products of 
  the initial conditions in the reaction, and therefore manage to clarify whether or not 
  the effects may be likely to become observable in the experiments. 
  One of the main features which can be an  observable in the VM is certainly the variation of the angular 
  distribution of the products. 
  In the case of proton transfer reactions between heavy $+$ light-heavy
  partners, such as $\rm{OH^{-} + C_{2}H_{2}}$~\cite{Farrar:06},  
  $\rm{NH_{3} + H_{2}O^{+}/H_{3}O^{+}}$~\cite{Yue:04,Farrar:04}, or  
  heavy $+$ light partners $\rm{Ar + H_{2}^{+}}$~\cite{Michaelsen:07}, 
  experimental scattering results reveal distinct forward scattering with little momentum transfer to 
  the ionic product. The above findings thus indicate the presence of a direct stripping mechanism 
  irrespective of the system being studied. On the contrary, for light $+$ light reactions 
  like, $\rm{D + H_{2}}$~\cite{Miller:89} or the ionic reaction considered here, 
  the theory suggests that a large amount of momentum transfer to the products may occur 
  so that backward scattering could be an important effect. Even if the reactive cross-sections can generally  
  be described by using a full quantum treatment, the simpler description of classically impulsive
  collisions can also help in some cases to qualitatively understand reaction mechanisms.

  In the following, we shall therefore  simulate the experimental velocity images of the products that result 
  from the exchange proton reaction between $\rm{He}$ and $\rm{H_{2}^{+}}$ which will be  considered 
  to be in different initial quantum states, aiming at identifying reaction mechanisms that could be directly 
  observed in the real imaging experiments which are currently being planned in our laboratory. As already 
  mentioned before, we shall include in the calculations the intrinsic aberration of our spectrometer in order to 
  underline more directly the links between our calculated DCSs and the experiments. 
   
  We shall use time-independent CC hyperspherical coordinate method to calculate the quantum mechanical 
  scattering matrix, from which, the state-to-state DCSs and the reaction probabilities are then calculated. 
  The above data are therefore employed to generate the fragment distributions in the velocity space 
  following the procedure that we shall discuss in the following sections.
  The calculations are based on a highly correlated reactive potential energy surface (PES) 
  computed with FCI/cc-pVQZ level of quantum chemical theory~\cite{Ram:09} which we shall also 
  briefly outline in the following section.

  This paper is structured as follows: In Section II, we provide an outline of the methodology that 
  we have employed to compute the time-independent Scattering Matrix and the ensuing angle-dependent 
  ingredients, i.e. the DCSs from which we shall construct the velocity maps. In the following Section III, we will 
  discuss the outcomes from our reactive quantum scattering calculations and then we will analyze 
  the results of our simulations once the angular distributions are presented in the format of the 
  experimentally observable velocity maps. The conclusions will be reported in the last Section IV.

\section{\label{sec:methods} Methods for calculations: an outline.}

\subsection{Quantum reactive scattering calculations}

In this work,  quantum scattering calculations were performed by using the standard version of the ABC code \cite{Skouteris:00}. 
The ABC program employs a time-independent coupled channel hyperspherical coordinate 
method to calculate the quantum mechanical scattering matrix. The set of coupled hyperradial 
equations is solved using a logarithmic derivative method. The output of the code are 
the parity-adapted $S$-matrix elements $S^{J,P}_{v'j'K', vjK} (E)$ which have to be converted into the 
standard helicity-representation $S$-matrix elements $S^{J}_{v'j'K', vjK} (E)$ \cite{Miller:89}. 

The observables of the reaction can be computed employing the helicity-representation $S$-matrix elements. 
The quantum $\rm{H_2}^{+}$$(v,j)$ +  He $\to$ $\rm{HHe}^{+}$$(v',j')$ + $\rm{H}$ state-to-state reactive differential cross sections 
are calculated as, 
\begin{eqnarray}
& & \displaystyle \frac{d \sigma_{v'j'K'  \leftarrow  vjK}}{d\Omega} (\theta ', E) \nonumber \\   
&&  = \left| \displaystyle \frac{1}{ 2ik_{vj}} \displaystyle \sum_{J } 
(2J+1) d_{K'K}^{J}(\theta ') S^{J }_{vjK,v'j'K'}(E) \right|^{2} , 
\end{eqnarray}
where $d_{K'K}^{J}(\theta ')$ are reduced rotational matrices \cite{Brink:68}, 
 $k_{vj}^{2} = 2 \mu E / \hbar^2$ and $\mu$ is the reduced mass of the initial fragments.
$J$ is the value of the total angular momentum quantum number. $K$ and $K'$ are the projection of the total 
angular momentum vector, $\mathbf{J}$, on the body-fixed 
$z$-axis of reactant and product Jacobi coordinates. 

The state-to-state integral reactive cross sections are then defined as  
\begin{eqnarray}
 \sigma_{v'j'K'  \leftarrow vjK}(E) && \equiv 2 \pi \displaystyle \int_{0}^{\pi} d\theta' sin(\theta ') 
\displaystyle \frac{ d \sigma_{vjK,v'j'K'}}{d\Omega} \nonumber \\   
&&  = \frac{\pi}{k^{2}_{vj}}   \displaystyle \sum_{J } 
(2J+1) \left|  S^{J}_{v'j'K',vjK} \right|^{2}  
\end{eqnarray}

The helicity averaged reaction probability is defined by 
\begin{eqnarray}\label{eq:ics}
P_{v'j' \leftarrow vj} (E) = (2j+1)^{-1} \displaystyle \sum_{J K K'} \left| S^{J }_{vjK,v'j'K'}(E) \right|^{2} 
\end{eqnarray}
One can also sum and average over the final and initial helicity
quantum numbers to determine the averaged reactive differential cross section, 
\begin{eqnarray}\label{eq:dcs}
\displaystyle \frac{ d \sigma_{j'v' \leftarrow jv}}{d \Omega}  (\theta ',E)
= (2j+1)^{-1}  \displaystyle \sum_{K K'} \displaystyle \frac{\sigma_{vjK,v'j'K'}}{d\Omega} (\theta ',E)
\end{eqnarray}
In the present and the following sections we shall omit the arrangement labels from all equations 
since everything will refer to a single reaction. It is also convenient to refer to the scattering angle of the 
molecular product in the center of mass as $\theta = \pi - \theta'$, 
and therefore we adopt this notation from this point onwards.

\subsection{\label{sec:vim}  Velocity map simulations}

In our simulation we analyze 30000 reactive collisions in which each fragment follows the probabilities dictated by equations
(\ref{eq:ics}) and (\ref{eq:dcs}). If the initial collision energy is well defined, then the 
final velocity of the ion $\rm{HeH^{+}}$$(j',v')$ is completely determined and it defines a sphere in the velocity 
space with the origin in the center of mass of the complex $\rm{HeH_{2}}^{+}$. 
The probability for the fragment to end on the surface of such sphere is determined by the state-to-state integral reaction 
probability given by eq.(\ref{eq:ics}).  
Then the orientation of the velocity vector is selected from the angular distributions provided by the 
state-to-state DCS (\ref{eq:dcs}).

Once the point is located in the VM, one still has to include the aberrations of the velocity mapping. 
To characterize the resolution of the electric field configuration, we calculate the radius of the impact position and the time on the detector by means of a Taylor expansion \cite{Wester:14}, 
\begin{equation}
X = \mathbf{D}_{X}^1|_{0} \mathbf{v} + \frac{1}{2} \mathbf{v} \mathbf{D}_{X}^2|_{0} \mathbf{v}^{T} + ...,
\end{equation}
where $X$ denotes either the spatial coordinate $R$ on the detector
surface or the time-of-flight to the detector $\tau$. The vector
$\mathbf{v} = (r,z,v_{r},v_{z})$ describes the ion 
position and velocity before the electric field of the VMI spectrometer is activated \cite{Chichinin:09}. 
The points of origin for $r$ and $R$ are located on the symmetry axis setup of our spectrometer \cite{Wester:14}. 
The origin for the $z$-direction is located in the middle between the two lowest plates of the setup, 
8 mm above the lowest one.
The time-of-flight $\tau$ is measured relative to the arrival time of a particle at rest 
that is starting at $r=0$ and $z=0$. $D_{X}^{1}(4\times1)$ and $D_{X}^{2}(4\times4)$ 
represent the respective 
first and second order matrices of partial derivatives. 
These matrices were computed for our VMI setup using numerical calculations with SIMION \cite{SIMION_brief}.
The matrix entries depend on the mass of the particle for which
they are determined.

In the simulations we have considered typical experimental conditions to compute the deviations 
of the points in the VM. For the proton transfer reaction considered here, the standard 
deviation of the initial ion energy was 0.1 eV, while the standard deviation of ion angle and the neutral 
angle are 1.0$^{\circ}$ and 4.1$^{\circ}$, respectively.

\subsection{\label{sec:pes} The reactive potential energy surface.}

Here we have used the most recent reactive PES computed for the present reactive system. The  details of 
its evaluation are described extensively in reference \cite{Ram:09}. The surface was obtained 
at very high level of quantum chemical calculations, using the MRCI approach that included all 
single and double excitations from the initial CASSCF space. The calculated ab initio 
points were fitted using the Aguado-Paniagua many-body expansion \cite{Aguado:98} in which the PES of the 
triatomic system ABC can be expressed as a sum of three monoatomic terms $(V^{1}_{A},V^{1}_{B},V^{1}_{C})$, 
three 2-body terms $(V^{2}_{AB},V^{2}_{BC},V^{2}_{AC})$ and one 3-body term $(V^{3}_{ABC})$. The latter term is 
defined as: 
\begin{equation}
V_{ABC}^{3}(R_{AB}, R_{AC}, R_{BC}) = \displaystyle \sum_{ijk}^{M} d_{ijk} \rho_{AB}^{i} \rho_{AC}^{j} \rho_{BC}^{k}
\end{equation}
where the optimal fitting was found by setting the parameter $M=8$ for those geometries at FCI/cc-pVQZ level 
of theory. Here $M$ is an order
index of the expansion of a product function that decays
exponentially with the distance and is defined in the reference. 
The fitting of the FCI/cc-pVQZ surface turns out to be very accurate and it has been  suggested as a promising potential for 
further dynamics studies \cite{Ram:09}.

\begin{figure} 
\includegraphics[width=8.cm, angle=270]{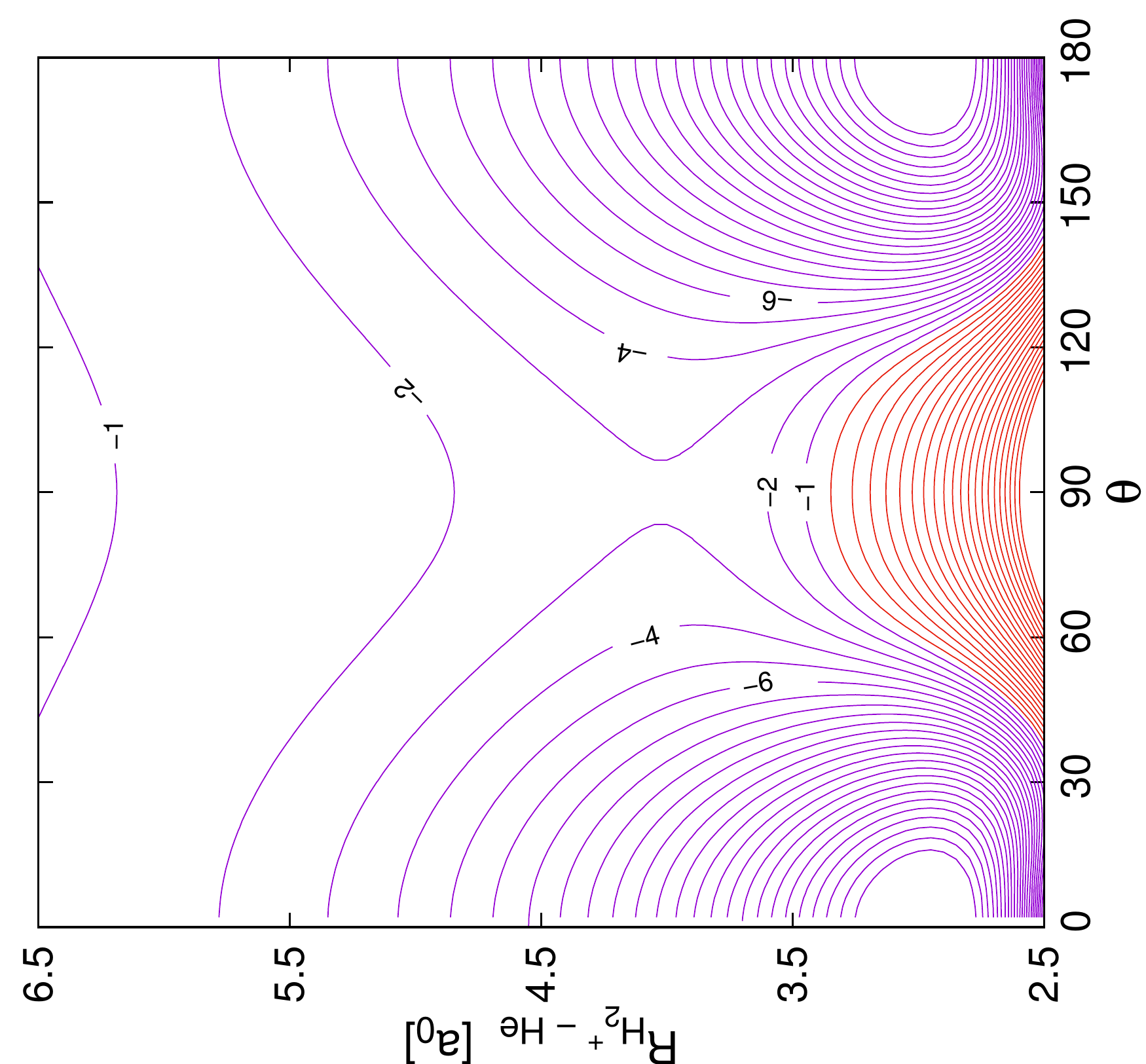} % Here is how to import EPS art
\includegraphics[width=8.cm, angle=270]{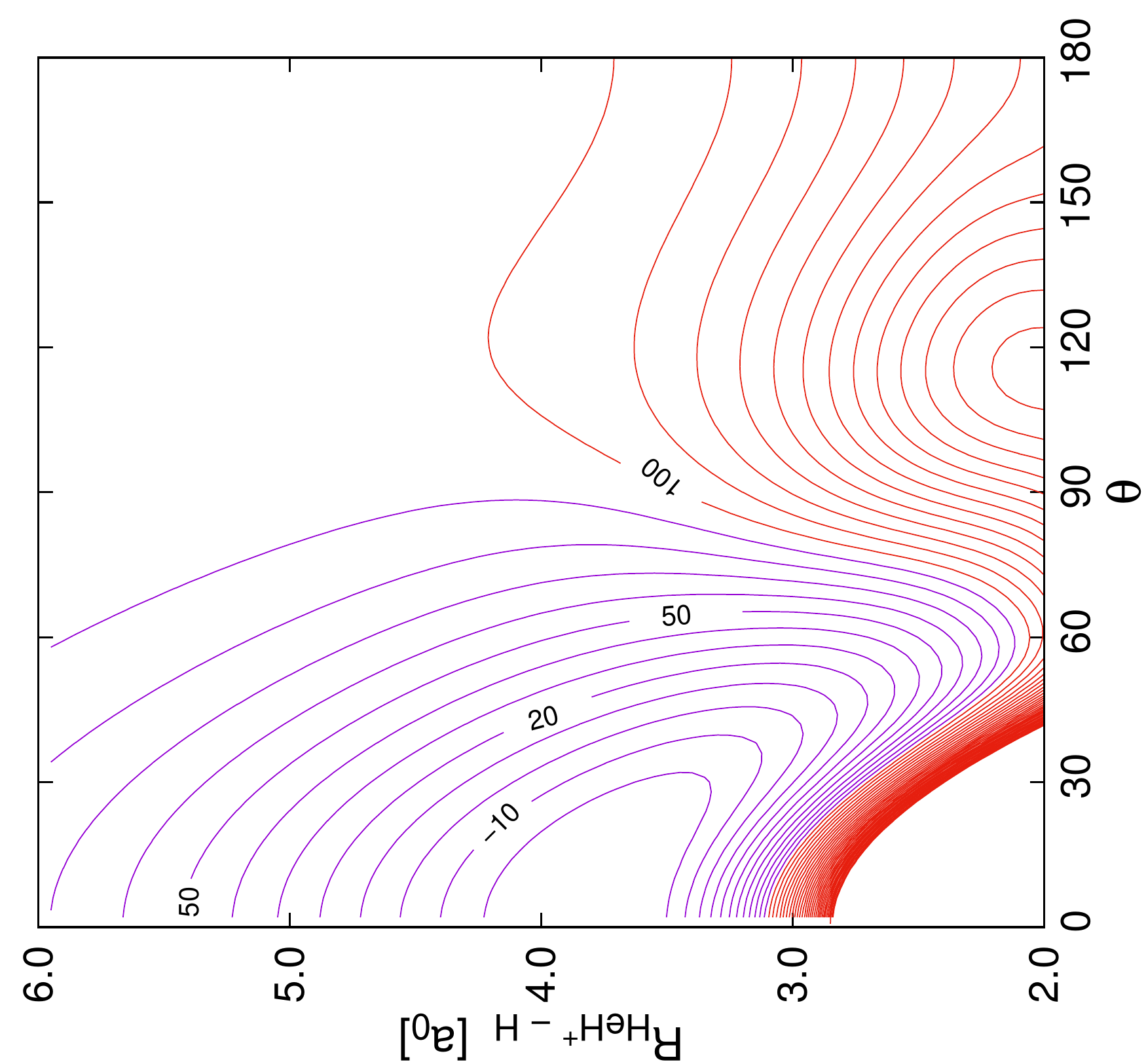} % Here is how to import EPS art
\caption{\label{fig:pes} Contour plot of the PES as a function of the Jacobi coordinates. 
Energies are in units of $10^2$ $\rm{cm}^{-1}$ and they are measured from the bottom of the asymptotic reactant valley.
(a) Contour plot of the reactants' PES constructed by taking the internuclear distance value of $\rm{H_{2}^{+}}$ at 2.074 au.  
(b) Contour plot of the products' PES constructed by taking the internuclear distance value of $\rm{HeH^{+}}$ at 1.927 au.  
In this condition, at $\theta=0$, the $\rm{H}$ fragment points toward the hydrogen atom of the $\rm{HeH}^{+}$ molecule. }
\end{figure}

The contour plots of the reactive PES as a function of the different Jacobi coordinates are shown in Fig.(\ref{fig:pes}) for reactants and products. The plots evidence 
the strong anisotropic difference between the interaction of the different fragments, suggesting 
that the most probable reaction pathway is characterized by linear geometrical configurations in 
both reactant and product spaces, with $\rm{H}$ pointing toward the hydrogen atom of $\rm{HeH^{+}}$ 
when the reaction fragments recede from each other.  
The PES shape in the reactant space clearly indicates that the
``abstraction'' mechanism that forms the $\rm{HeH^{+}}$ on the
external region of the rotating $\rm{H_{2}}^{+}$ should be the predominant one in the reaction, as it has also been  
demonstrated for the inverse reaction \cite{Bovino:12}.

\begin{figure} 
%\vspace{5cm}
\includegraphics[width=8.cm, angle=0]{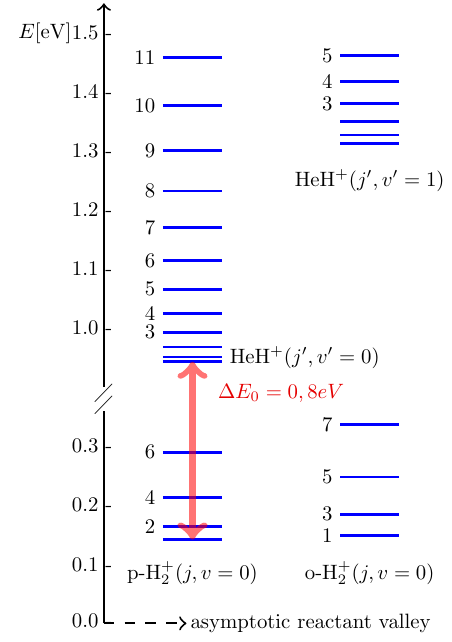} % Here is how to import EPS art
\caption{\label{fig:energy_levels} Schematic description of some energy levels of the reactants and products for the  
$\rm{He}$ + $\rm{H}_{2}^{+}$$(v,j)$ $\to \rm{HeH}^{+}$$(j',v')$ + $\rm{H}$ endoergic reaction.  }
\end{figure}

The Reaction (\ref{eq:reaction}) turns out to be an endoergic reaction that is quickly promoted by increasing the collision energy  
above the threshold. 
We show in Figure \ref{fig:energy_levels} the  endoergicity of the reactions and all the 
ro-vibrational levels of $\rm{HeH^{+}(j',v')}$ that are opened  
at 1.5 eV, the total collision energy used in our simulations. 
The initial ro-vibrational energy levels considered in this work for the reactants p-$\rm{H}_{2}^{+}$ and o-$\rm{H}_{2}^{+}$
are also depicted on that figure. 
We study only reactions that start with the reactants in the lower excited 
rotational states, e.g. $j \leq 7$ and $v=0$. In this situation, the initial kinetic energy of the fragments supplies most of the 
energy for breaking the strong $\rm{H}_{2}^{+}$ bound and for exciting the internal modes of $\rm{HeH}^{+}$. 
The above considerations 
readily tell us that we can computationally generate the angular distributions of the product molecules 
by starting with different rotational states of the $\rm{H_{2}^+}$ and therefore observe from the final patterns 
of such angular distributions possible effects from the internal states of the molecular partners on 
the outcomes of the reaction, when such internal energy can end up into different rovibrational states 
of the product molecular ion.

\section{Computational results: a link to the experiments}

In this section we intend to analyze  in some detail reaction (\ref{eq:reaction}) in order 
to unravel its nanoscopic behavior from a computational study of its reactive DCS 
and from linking such dynamical observables with VM simulations obtained from 
possible specific experiments.

\begin{table}
\caption{\label{tab:table1} Total ICS for the reaction $\rm{He}$ + $\rm{H_{2}}^{+}$$(j,v)$ $\to \rm{HeH}^{+}$ + $\rm{H}$  
as a function of the translational collision energy. Our present results reproduce very closely 
the state of the art calculations reported in reference \cite{Fazio:12}
and they are in a good agreement with the recent experimental results \cite{Tang:02}. }
\begin{ruledtabular}
\begin{tabular}{lll   llllllll }
$(v,j)$   &  $E_{c}$ [eV]          & $\sigma( vj )$ [{\normalfont\AA}$^{2}$] (this work) & $\sigma( vj )$ [{\normalfont\AA}$^{2}$]  \\
\hline
0,0    & 1.357      & 1.779(-2)  & 1.776(-2)\footnote{ICS from reference [7] and computed with the RFCI PES (the same PES used in the present work)}  \\
1,0              & 1.085      & 0.384      & 0.385$\rm{^{a}}$       \\
2,0              & 0.829      & 2.051      & 2.051$\rm{^{a}}$       \\
3,0              & 0.588      & 4.003      & 4.008$\rm{^{a}}$        \\
0,1    & 1.350      & 1.830(-2)  & 1.909(-2)\footnote{ICS from reference [7] and computed with the RMRCI PES}   \\
1,1              & 1.079      & 0.384      & 0.387$\rm{^{b}}$        \\
2,1              & 0.823      & 2.019      & 2.012$\rm{^{b}}$       \\
3,1              & 0.582      & 3.941      & 3.975$\rm{^{b}}$      \\
4,1              & 0.355      & 5.035      & 5.016$\rm{^{b}}$       \\
\end{tabular}
\end{ruledtabular}
\end{table}

We have computed the $S$-matrix of the reaction using the accurate PES presented in Section \ref{sec:pes} 
and employing the same parameter used in the benchmark work of the De Fazio et. al \cite{Fazio:12}
However, in contrast with the above mentioned work that analysed the integral cross sections of the reaction, 
here we focus on the computation of the DCSs and its application in simulating image experiments.
The scattering calculations were carried out with the ABC program \cite{Skouteris:00} and 
the input parameters we have employed at 1.2 and 1.5 eV of total collision energies are the following: 
the total angular momentum 
quantum number is $J_{totmax} = 54$, the maximum values of the the helicity 
quantum number is $K_{max}= 12$, the maximum rotational quantum number of any channel is $j_{max}=26$, the 
maximum hyperradius is $\rho_{max}= 15 $ a$_0$, and the internal energy in any channel is $e_{max}=2.3$ eV. 
The selection of the above total energy values for the present calculations was suggested by the 
desire to run our numerical modeling under typical conditions of future experimental observations, whereby the 
endothermic reaction is fully energetically accessible and several cases of different reagents' and products' internal 
energy distributions could be analyzed.

Our calculations of the total integral reactive cross sections are in excellent agreement with reference \cite{Fazio:12} and 
with one of the most recent experiment \cite{Tang:02}. Some of our ICS are presented in table \ref{tab:table1}; they basically 
reproduce some of the ICSs presented in Figure 2 and 5 of reference \cite{Fazio:12}. 
Those authors indicate in their work that they have found better agreement with the existing experiments
results for reactions that occur between He and the ion $\rm{H_{2}^{+}}$$(v,j=1)$ in its  first $v=1$ and second $v=2$
vibrational excited states. 
However, the disagreement found by the calculations of De Fazio et. al\cite{Fazio:12} with the existing experiments 
for reactions promoted  by  $\rm{H_{2}^{+}}$$(v=0,j=1)$ in its ground vibrational state  
is not as yet completely clear and it may 
be indicative of uncertainties in the available experiments rather than with problems on the accuracy of
the reactive PES.  In this work we shall then focus on the reactions between $\rm{He}$ and the ion $\rm{H_{2}^{+}}$ ideally prepared in a well defined initial rotational ($j < 7 $) and vibrational states ($v=0$ or $v=1$).

\begin{figure}
\hspace*{5cm}
\includegraphics[width=10.cm, angle=0]{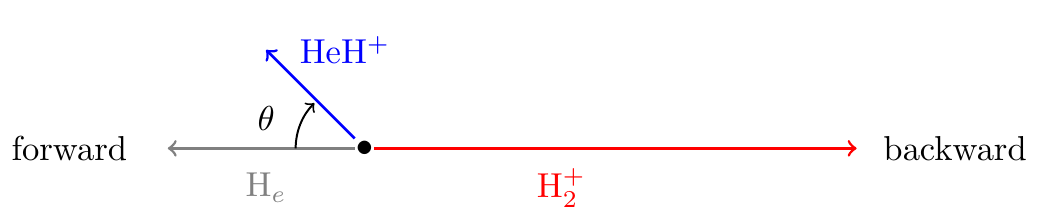} \\ % Here is how to import EPS art
\hspace*{-4cm}
%\vspace{-1cm}
\includegraphics[width=19.cm, angle=0]{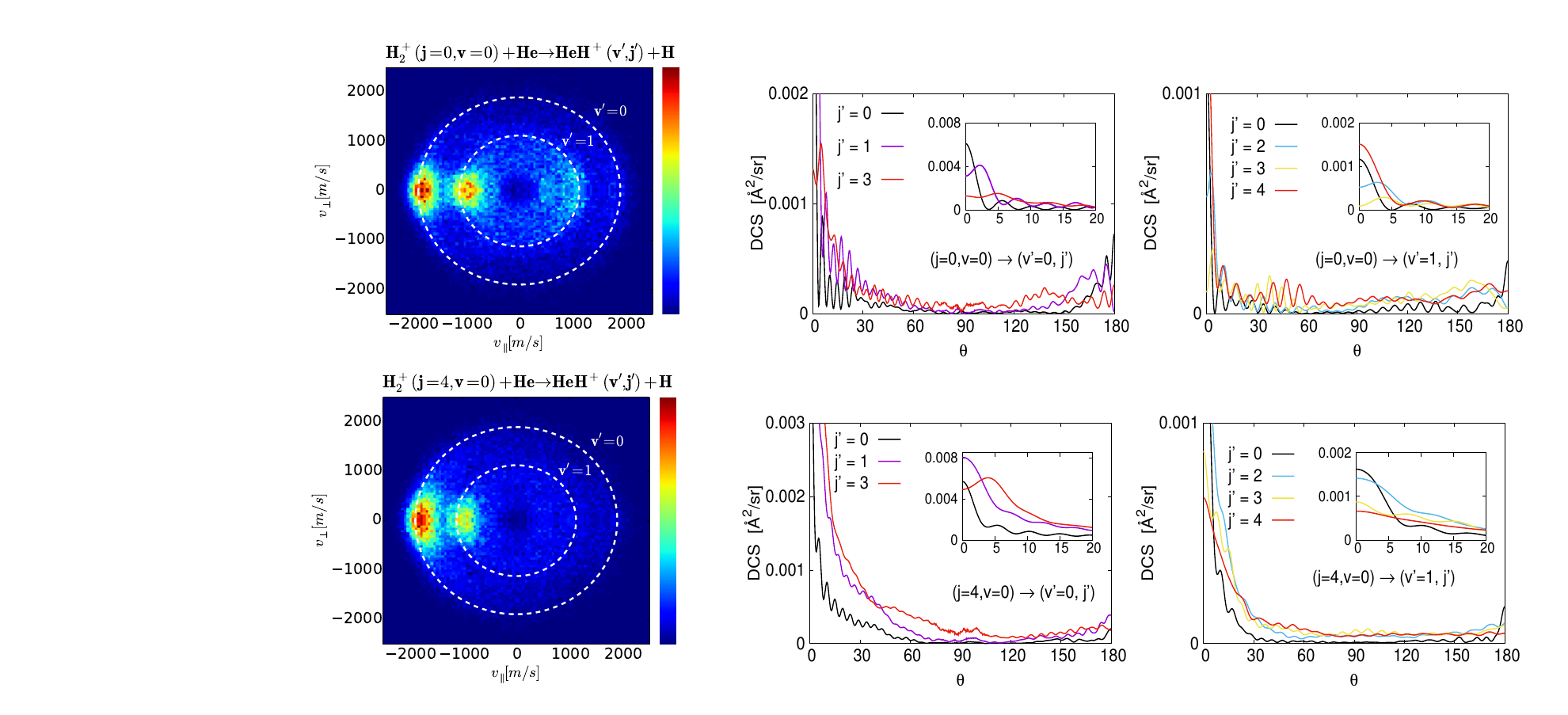} \\% Here is how to import EPS art
\hspace*{-4.0cm}
\includegraphics[width=19.cm, angle=0]{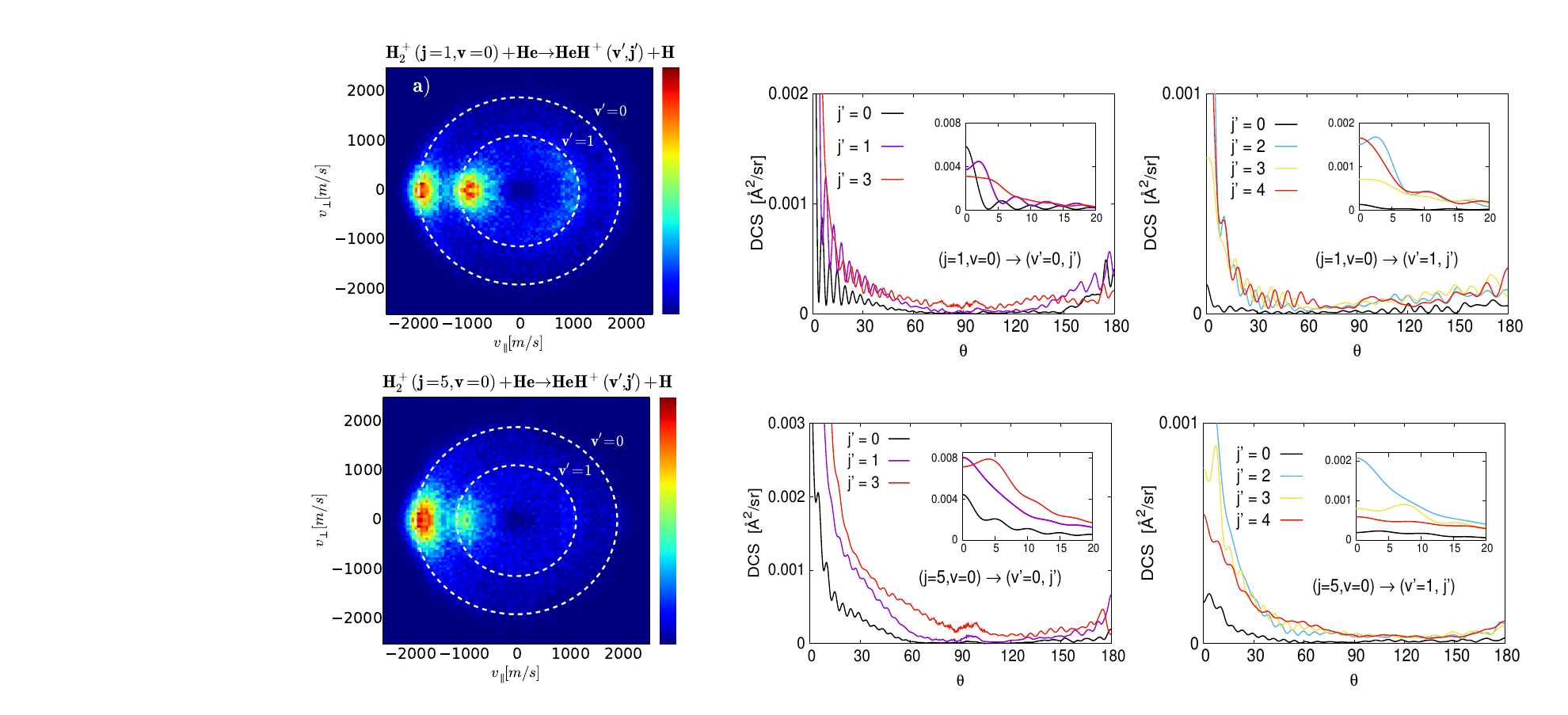}% Here is how to import EPS art
\caption{  \label{fig:vm} Velocity map simulation of the experiments and computed DCS for the title reaction 
at 1.5 eV of total collision energy.  
The eight panels on the right show some DCSs results while the four panels on the left report the different 
velocity maps obtained from the same $S$-matrix calculations. The top panel in this figure reports the products orientation 
in the velocity map frames.  }
\end{figure}

Following the methods outlined in the previous Section \ref{sec:methods}, we have computed the DCSs 
for different initial states (and hence relative collision energies) of the reacting molecule 
and focusing on different internal states of the product polar molecular ion. 
Some of the results we have 
obtained are given in the panels of Figure \ref{fig:vm}. 
The following considerations could be a perusal of the many results reported 
by the panels of Figure \ref{fig:vm}:

(i) All the DCSs clearly indicate that the reactions have a higher probability of occurring 
(larger computed values) when the product molecule is obtained in the lower excited rotational 
states of its ground vibrational level.  
In fact, the calculations show that there is a clear propensity to produce ions in the lower rotational states with $j' < 5$,
while the formation of $\rm{HeH}^{+}$ in a vibrationally excited state is visibly associated with 
lower reaction probabilities.

(ii) The corresponding velocity maps indicate an additional effect from the reaction dynamics: when molecular 
products are obtained in the first rotational levels of both $v'=0$ and $v'=1$ vibrational states,
the reaction is largely producing  them in the forward direction. 
However, when the final ionic product is obtained rotationally excited 
with $j' > 9$, we found that the backward scattering is promoted.

(iii) The computed angular distributions further show series of rapid oscillations superimposed on the 
decreasing values of the DCSs as the scattering angle increases. One further sees clearly that such oscillations 
become much less marked, and with reduced frequencies, when the ionic molecular product is obtained in increasingly 
higher rotational states.

(iv) The velocity maps also show us that the dominance of the forward scattering peaks is somewhat reduced 
when the reacting molecular ion initiates the process in an excited rotational level. The data of such maps reported 
in the 2nd and 4th row of the figure (left-hand-side) indicate that the products are obtained over a 
larger angular range than when the initial reacting partner molecule is in its lowest rotational state. 
We shall further discuss this aspect later in this section.

\begin{figure}
\includegraphics[width=8cm, angle=0]{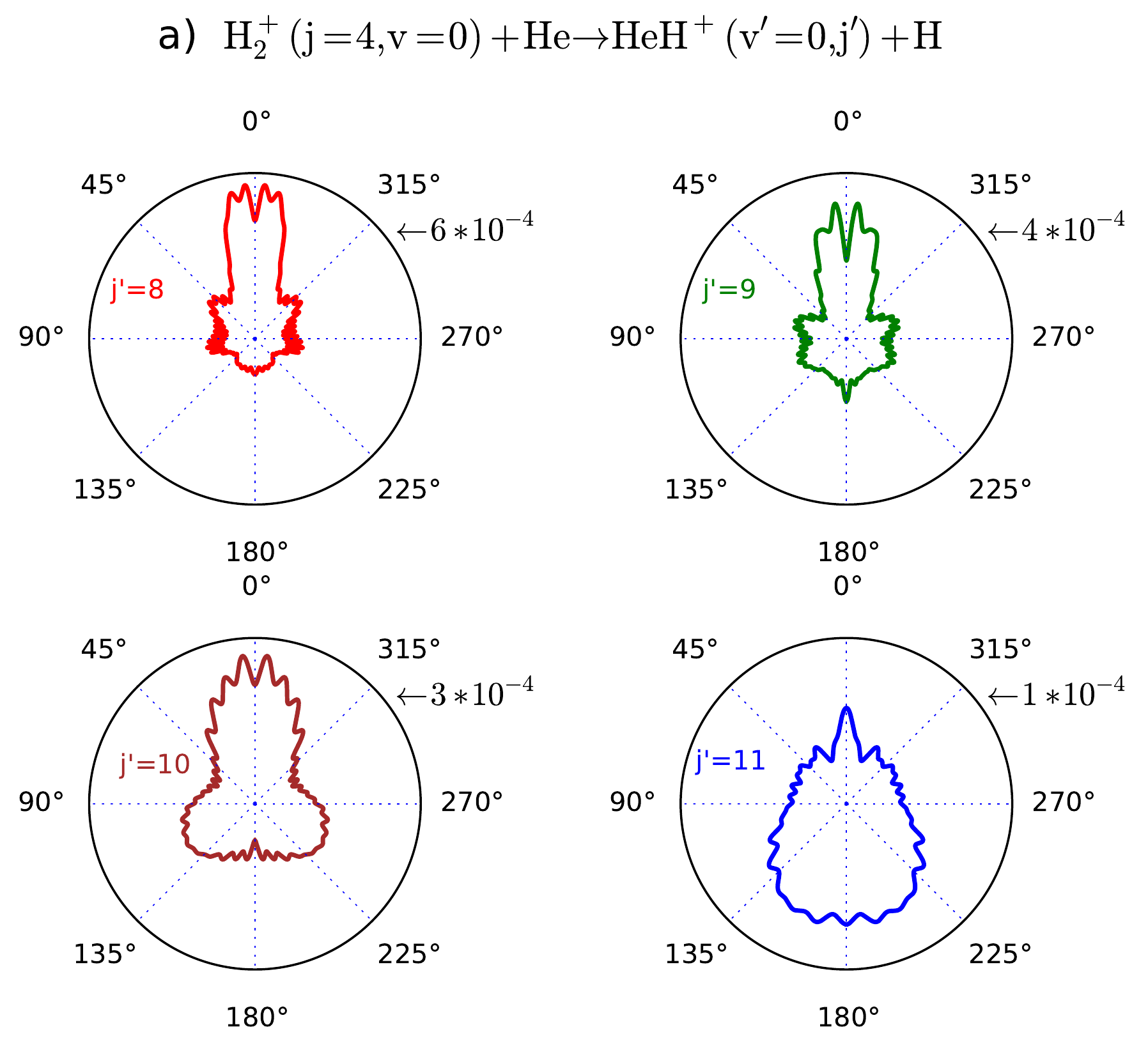}  % Here is how to import EPS art
\hspace*{0.1cm}
\includegraphics[width=8cm, angle=0]{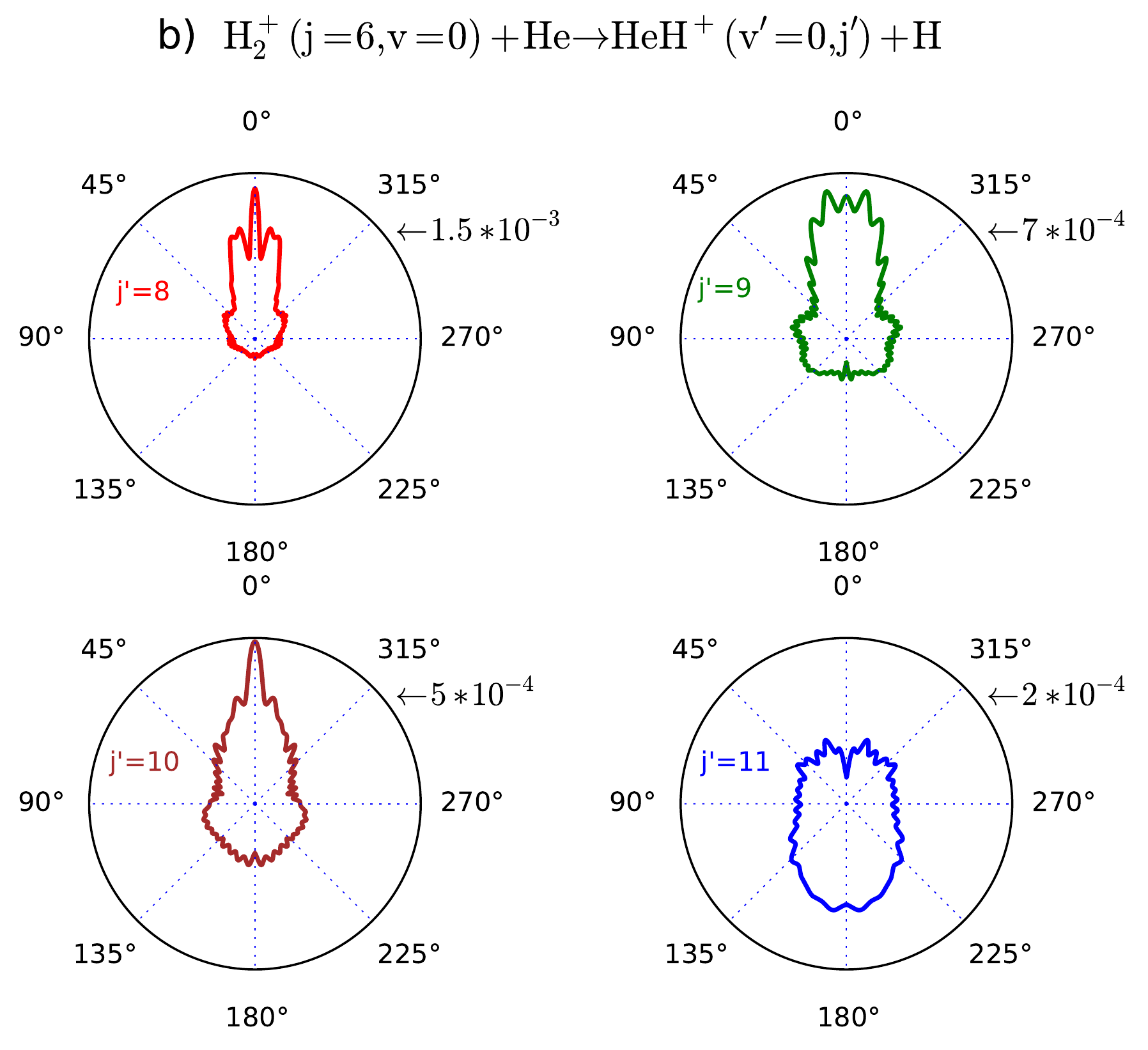} \\  % Here is how to import EPS art
\vspace{0.8cm}
\includegraphics[width=8cm, angle=0]{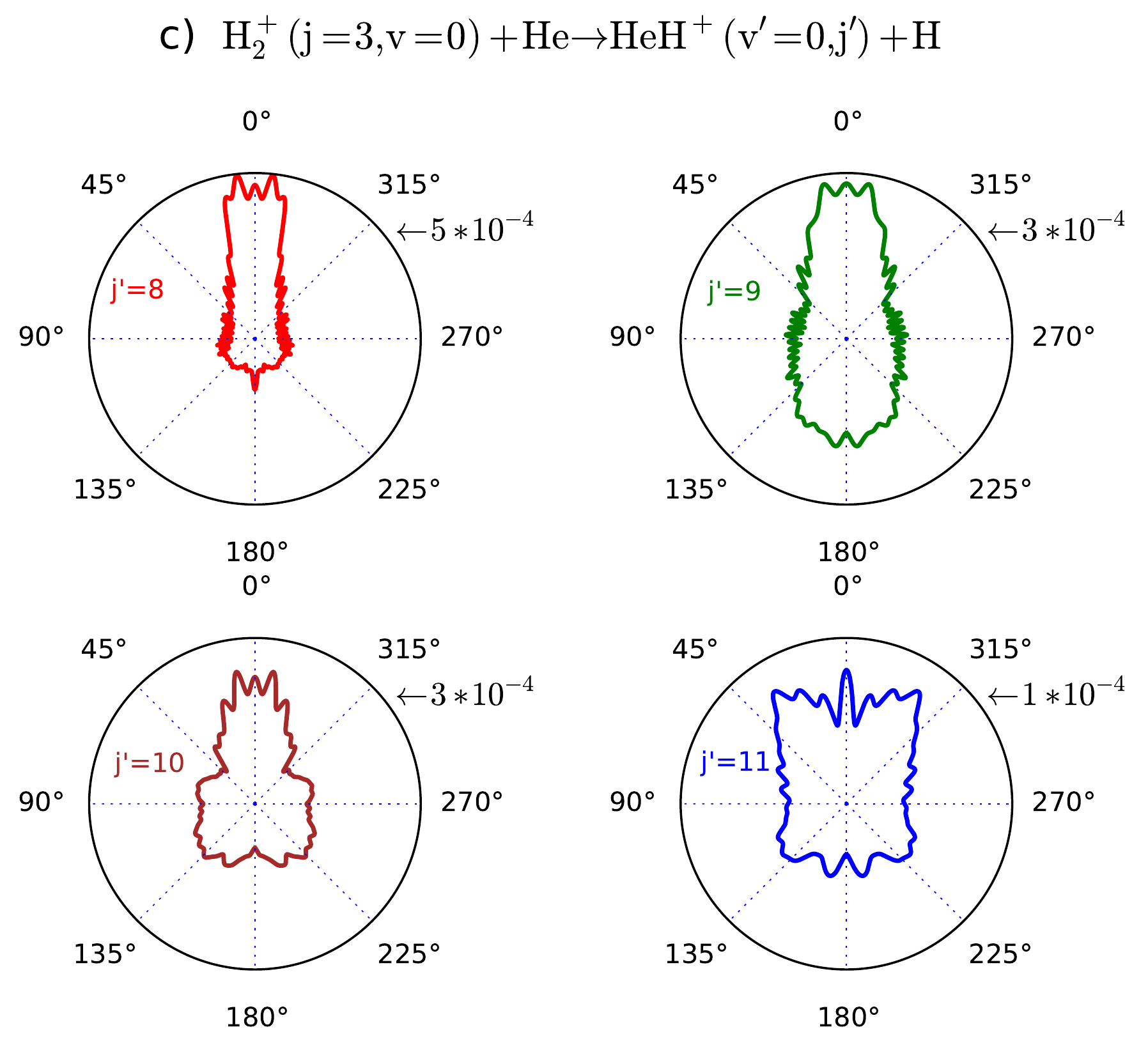}  % Here is how to import EPS art
\hspace*{0.1cm}
\includegraphics[width=8cm, angle=0]{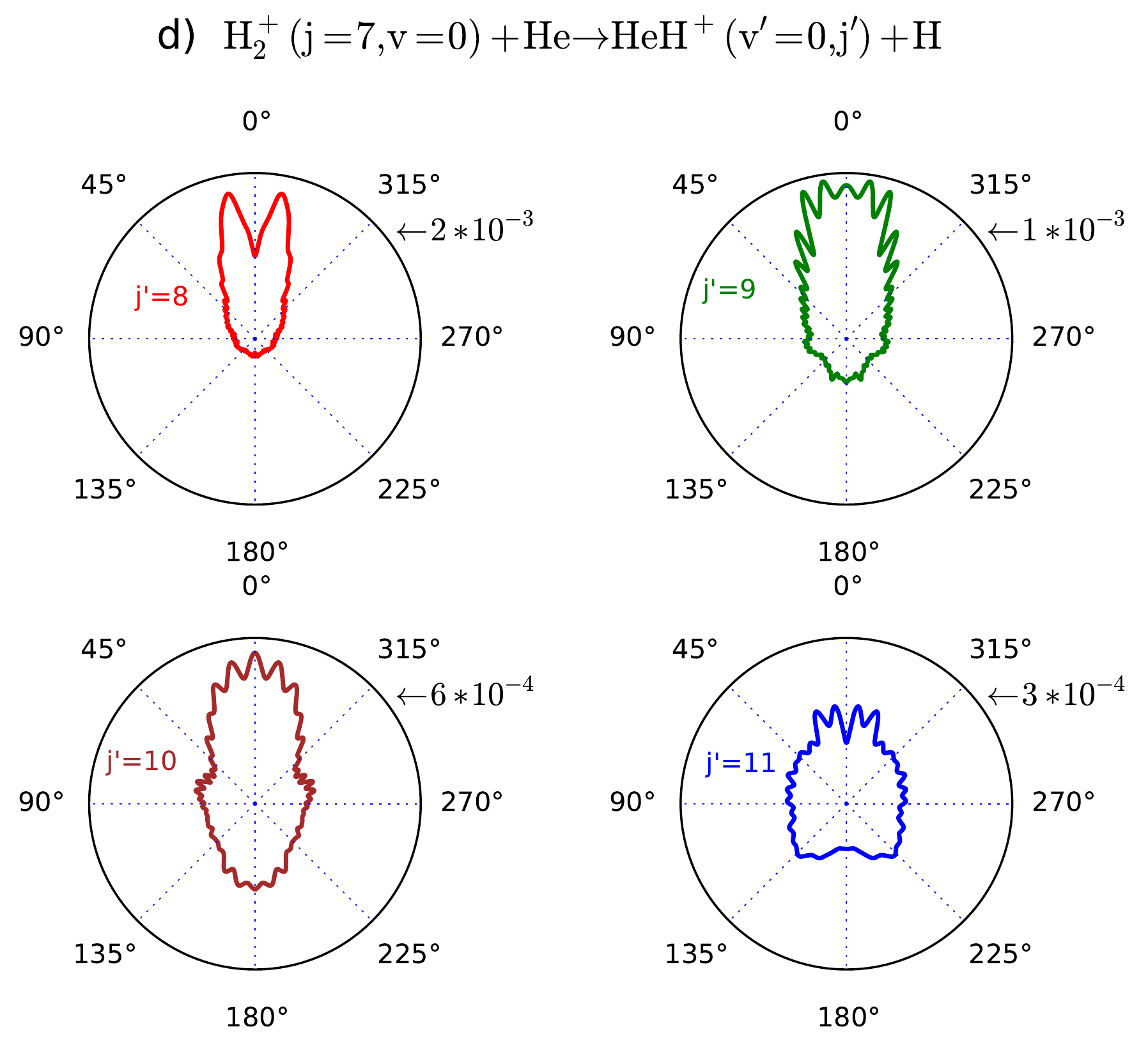}  % Here is how to import EPS art
%\vspace*{2cm}
\caption{\label{fig:polar}  Polar angular presentation of the product angular distributions for reactions involving 
p-$\rm{H_2}^+$ (panels (a) and (b)) and o-$\rm{H_2}^+$ (panels (c) and (d)). The four circular maps in each panel correspond 
to different final rotational states of the polar molecular product of the present reaction, increasing 
from left to right and from top to bottom. Each circle in all the panels indicate an upper limit for the specific intensity 
value of the DCSs for that process. See main text for further details. }
\end{figure}

With the intent of unraveling even more the interplay of the molecular internal states with 
the interaction potential of this system that guides the reactive process analyzed in this work, 
we report  a different presentation of the angular distributions obtained from our calculations in Figure \ref{fig:polar}.

The top panels of Figure \ref{fig:polar} indicate the angular distributions of polar molecular products 
as the $\rm{HeH}^+$ ions are produced in increasingly higher rotational states and when they originate with reactions 
involving p-$\rm{H}_{2}^+$ partners. The latter are initially in the $j=4$ state in panel a) while 
they are in their $j=6$ state in panel b). 
At least two clear features can be observed from the angular distributions reported in these panels:

(i) The ejection of the products into the backward direction is gradually enhanced 
as the ionic molecule is produced with higher values of $j'$. The highest 
rotational state available at the present collision energy ($j' = 11$) show a dominant ejection 
of the products into the backward direction.

(ii) all the circular maps in the two panels indicate how sensitive the reaction mechanisms are to the initial 
quantum states of p-$\rm{H}_{2}^+$: the maximum values of the DCSs may increase more than a factor of 2 
in some directions, while their angular dependence could also become completely altered.

The first of the observed features indicated above can be understood by looking at the relative energy levels reported 
by Figure \ref{fig:energy_levels}. The high initial kinetic energy is quickly consumed in the process of breaking the 
strong chemical bond of $\rm{H_{2}}^{+}$; this effect is favored by a ``billiard'' effect for this homonuclear molecule. 
Then if the surplus of energy gets channeled  toward the excitation of high rotational states ($j' > 7$) the relative 
kinetic energy of both product fragments is considerably reduced by about 0.4 eV and therefore the time allowed for the 
fragments to interact is increased. 
In addition, there is a second effect: when $j'>7$, higher values of the relative angular momentum ($l'$) 
are included in the CC equations.  When $l'>10$ the shape of the ensuing repulsive barrier can make it  
even more difficult for the product fragments to escape after entering the reaction region, thereby 
increasing the time of interaction and therefore 
the angular deflection of the product fragments. 
The combination of these effects with the strong anisotropic features of the products' potential 
interaction (see Figure \ref{fig:pes}) can therefore strongly alter 
the balance between backward and forward symmetry of products.

The analysis of the reactions involving o-$\rm{H}^{+}_{2}$ as the the reacting molecular 
partner (Figure \ref{fig:polar}.c and \ref{fig:polar}.d) indicates that          
the same arguments used for collisions involving p-$\rm{H}^{+}_{2}$ can be used in this case as well. 
However, the explanation of the specific  differences observed for each state-to-state DCSs of Figure \ref{fig:polar}
is more complex. The specific differences between state-to-state non-reactive,  inelastic DCSs have been   
explained by the ``Fraunhofer theory'' \cite{Faubel:84}. 
For reactive collisions, these effects may also be introduced by using statistical treatment based 
on correlation theory \cite{Herschbach:75}. The possible application of the above approach 
will however be attempted in a future, separate publication.

\begin{figure} 
%\vspace{-2cm}
\includegraphics[width=8cm, angle=0]{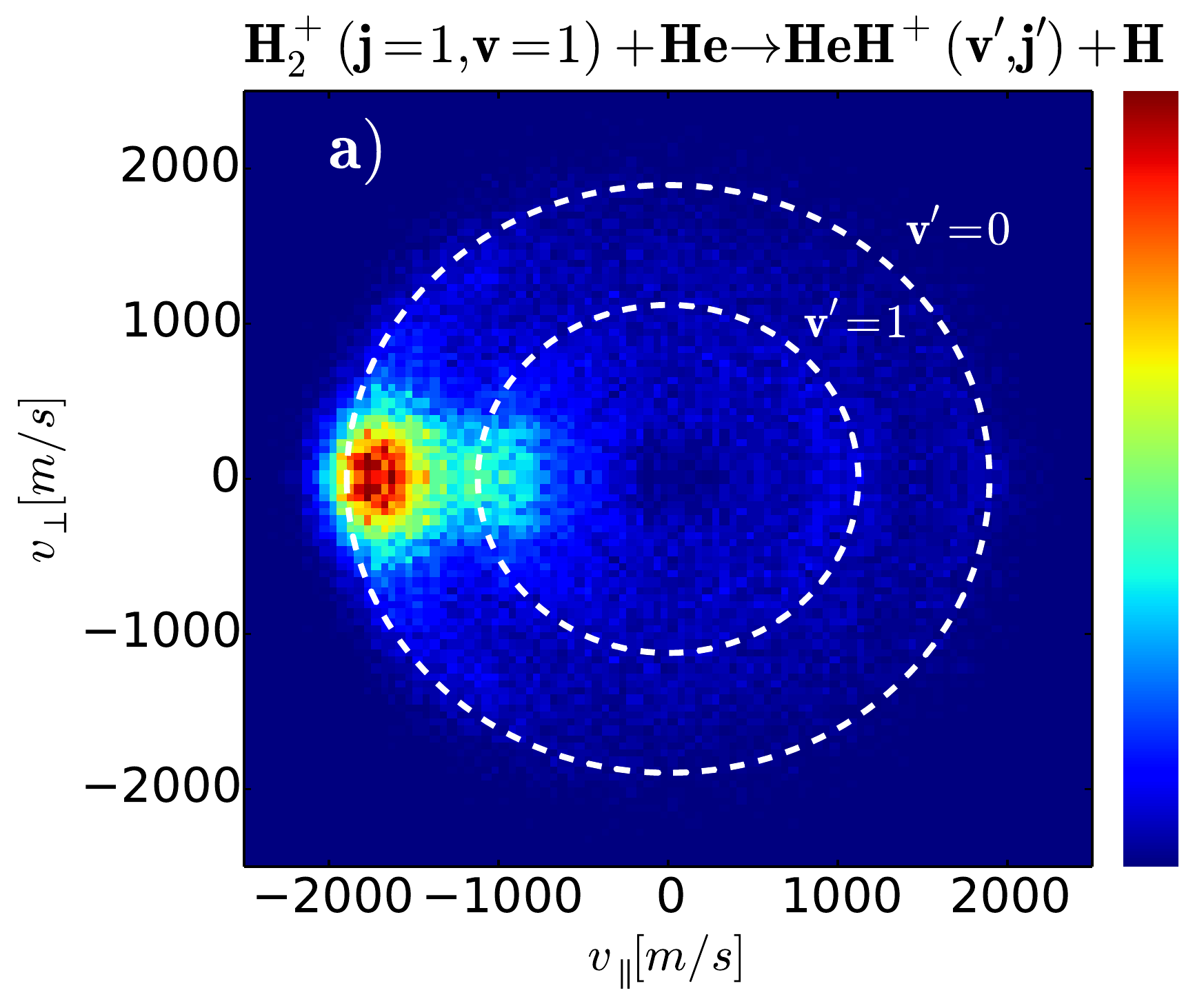}  % Here is how to import EPS art
\includegraphics[width=8cm, angle=0]{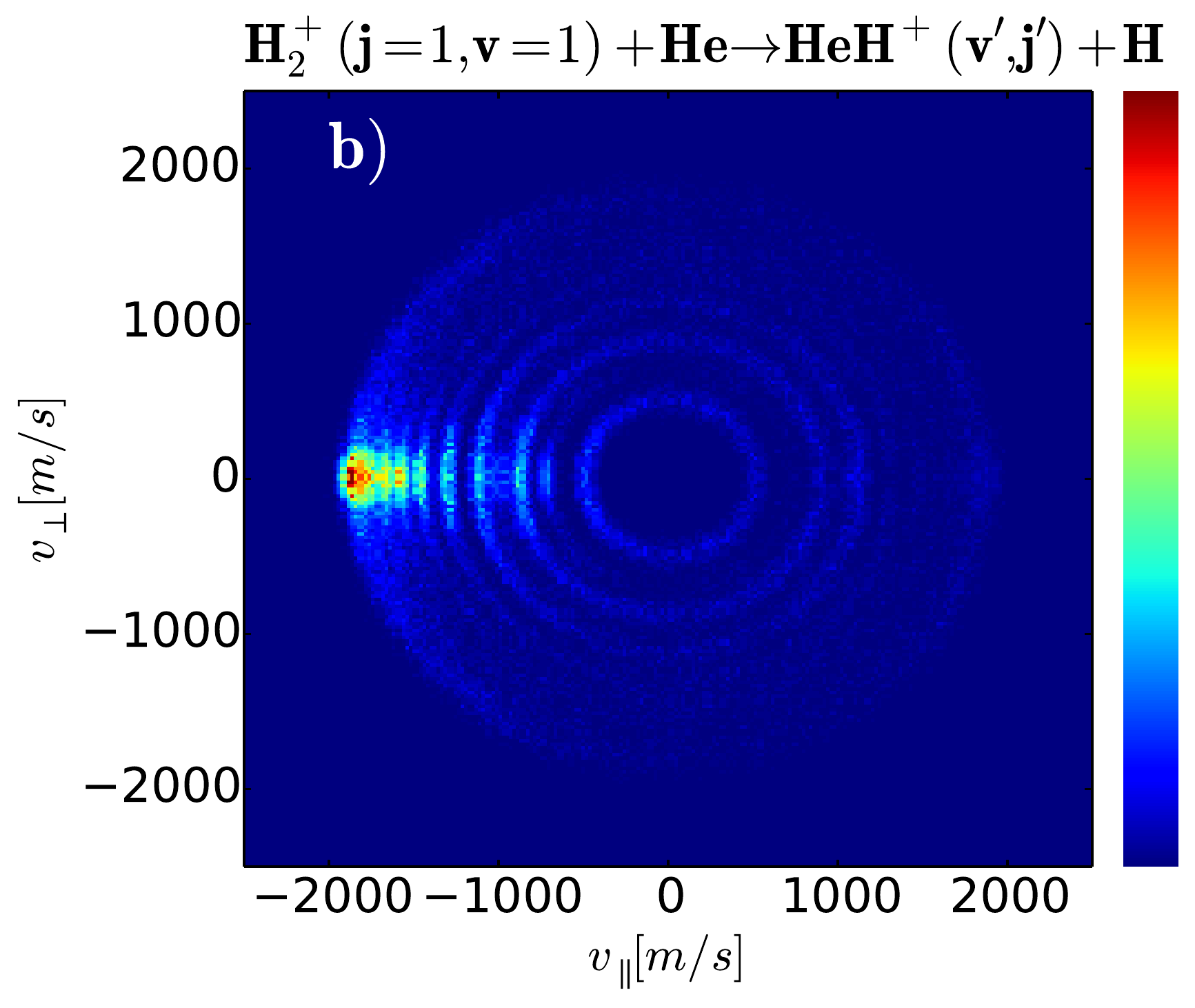} \\   % Here is how to import EPS art
\includegraphics[width=8cm, angle=0]{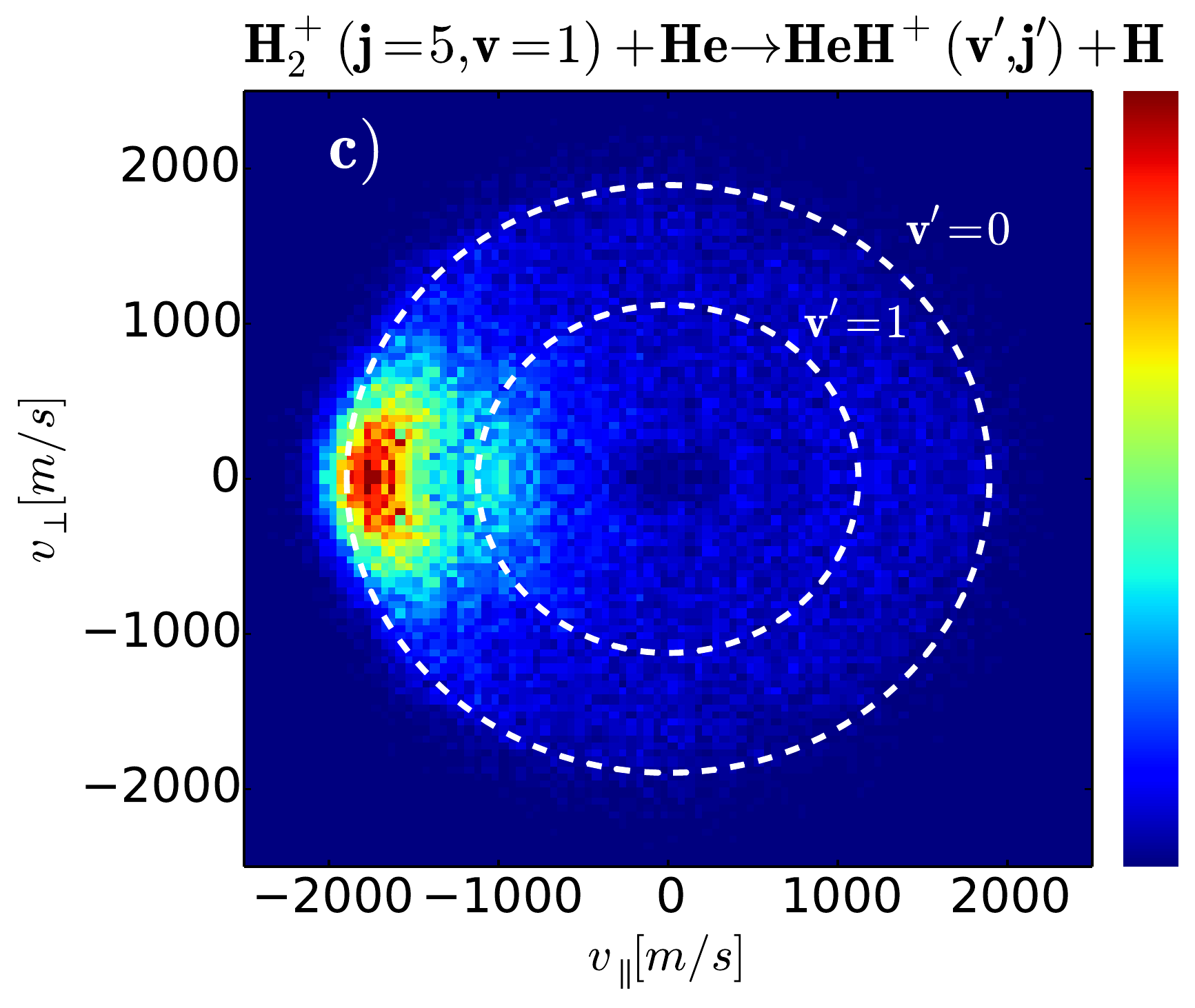}  % Here is how to import EPS art
\includegraphics[width=8cm, angle=0]{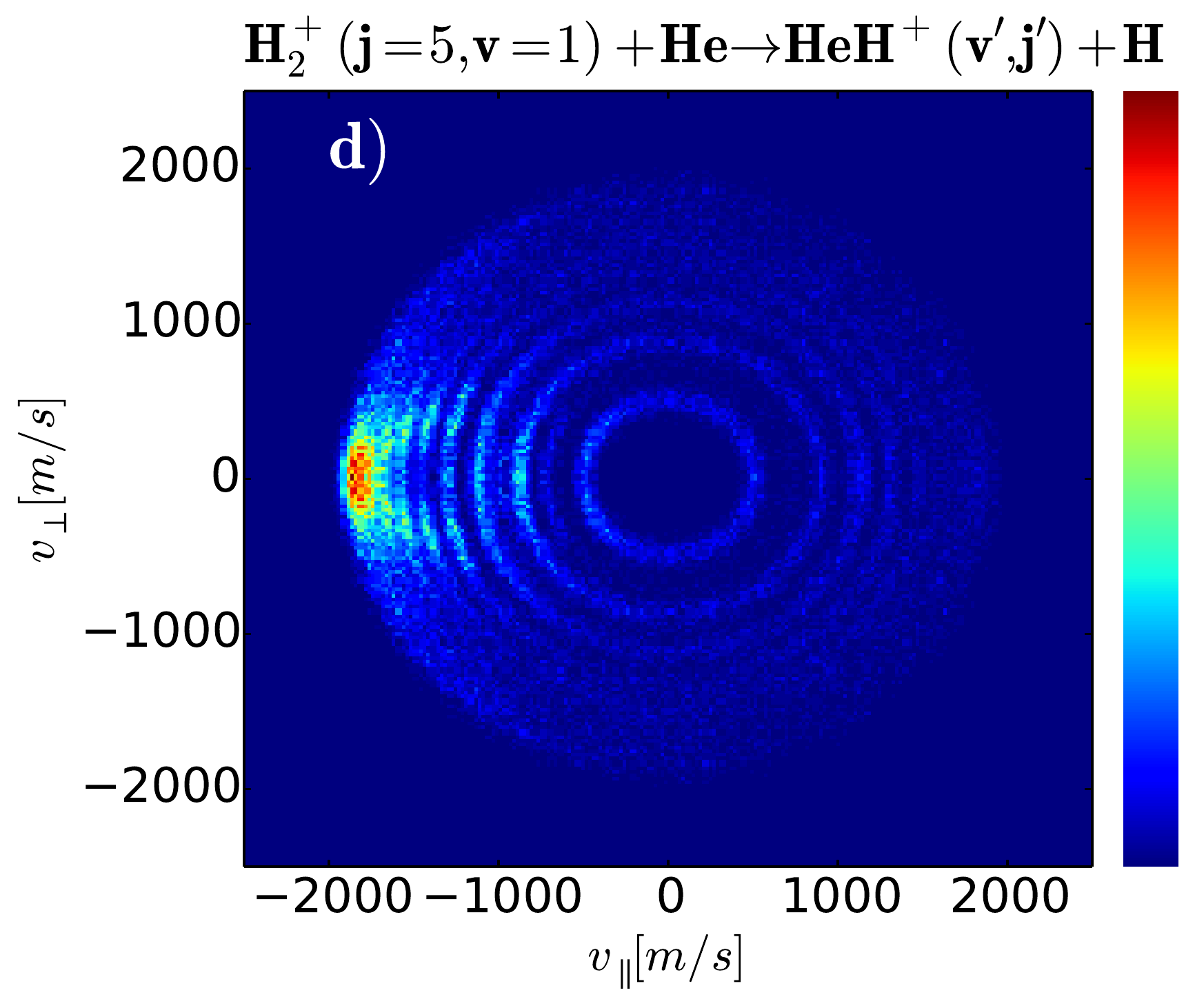}  % Here is how to import EPS art 
% \vspace{3cm}
\caption{\label{fig:vm2} Simulation of the VM images experiments for the title reaction. The upper panels show 
the reaction between  o-$\rm{H}_{2}^{+}(v=1,j=1)$ and $\rm{He}$ occurring at 1.08 eV of collision energy. 
The lower panels show the reaction between  o-$\rm{H}_{2}^{+}(v=1,j=5)$ and $\rm{He}$ occurring at 0.984 eV.
In order to disentangle the real simulation presented in Figures a) and c), we display in Figures b) and d)  
a non-realistic simulation for which the actual standard deviation of every point in the VM is reduced by a factor of 4. }
\end{figure}

Is it also possible to distinguish among experimental images of reactive collisions between 
ions prepared in the first initial vibrational excited state $v=1$ but at different 
rotational quantum states (see Figure \ref{fig:vm2}). 
In this situation, the main effect is that the images broaden toward larger values of $\theta$ around the forward direction 
as $j$ increases. The differences between the images is more evident in the region close to the the kinematical 
cut-off where the maximum density of rotational levels for $v' = 0$ are localized. 
The non-realistic (ideal experiments) simulations presented on the right hand side of Figure \ref{fig:vm2}, 
reveal this effect with more clarity. 
For reactions that begin with ions in $j=1$, the angular distribution around $\theta=0$ shrinks as $j'$ increases, 
while the contrary effect is observed for the reactions that start with the reactants in $j=5$.  
The above effect suggests a more effective transfer of angular momentum in reactions that are promoted by the 
$\rm{H_{2}}^{+}$ ions that are already rotationally excited.

There are also variations between images of reactions that occur with the  $\rm{H_{2}}^{+}$ ions in 
different initial vibrational quantum states but in the same 
initial rotational states. 
If one compares, in fact, the reactions between o-$\rm{H_{2}}^{+}(j=1,v)$ and $\rm{He}$ depicted in the images of 
Figures \ref{fig:vm} and \ref{fig:vm2}, it is possible to observe that the first excited vibrational 
levels of the products $\rm{HeH}^{+}$ 
is more easily populated by reactants prepared in the ground vibrational state $v=0$.  
Moreover, reactants in the state $\ket{v=0,j=1}$ promote more backward scattering than those in the state $\ket{v=1,j=1}$.
The above finding suggests that the capture of one hydrogen atom by He in the forward direction 
is less  favoured when $\rm{H}_{2}^{+}$ is in its most stable  
ground vibrational state. 
However, the capture mechanism in the forward direction is enhanced by the vibration of $\rm{H}_{2}^{+}$ that  
increase the size of the ``obstacle'' experienced by $\rm{He}$ and exposes more to the reaction those  
hydrogen atoms which are away from the molecular equilibrium position. 
A more rigorous description of this mechanism, however,  should take into account a more extensive analysis of the specific features of the reactive PES, the strength of the  rotational coupling and the rotational correlations that occur during the reaction.

\begin{figure} 
%\vspace{3cm}
%\hspace{-5cm}
\includegraphics[width=7cm, angle=0]{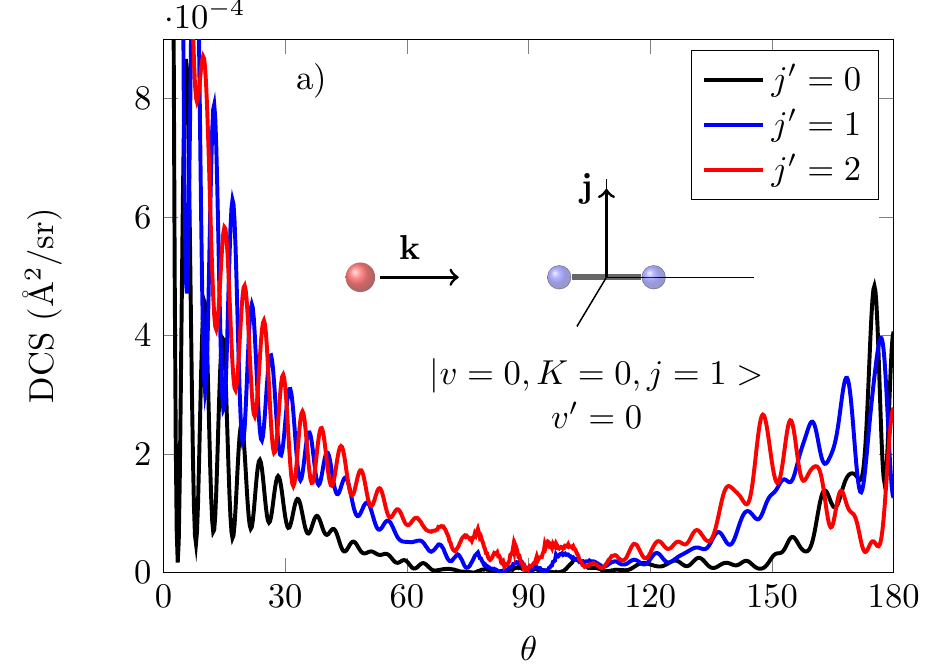}   % Here is how to import EPS art
\hspace{2cm}
\includegraphics[width=7cm, angle=0]{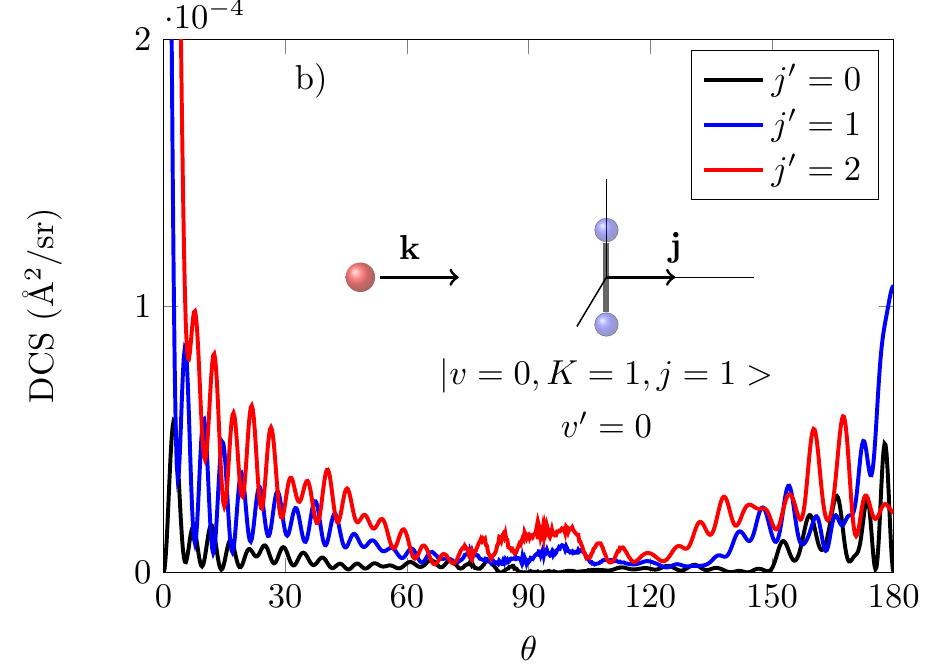} \\  % Here is how to import EPS art
%\vspace{3cm}
%\hspace{-5cm}
\includegraphics[width=7cm, angle=0]{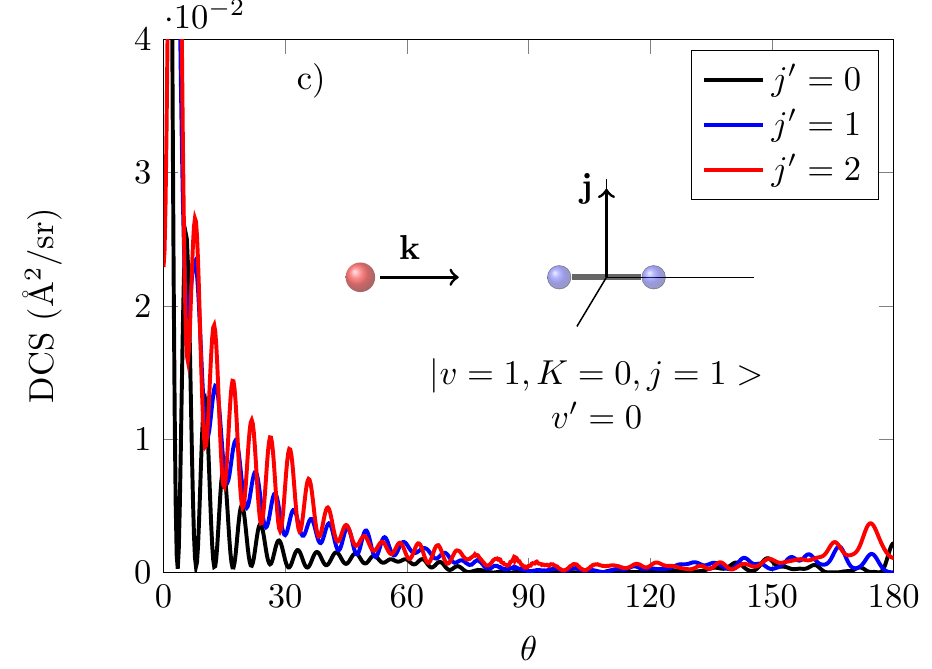} % Here is how to import EPS art
\hspace{2cm}
\includegraphics[width=7cm, angle=0]{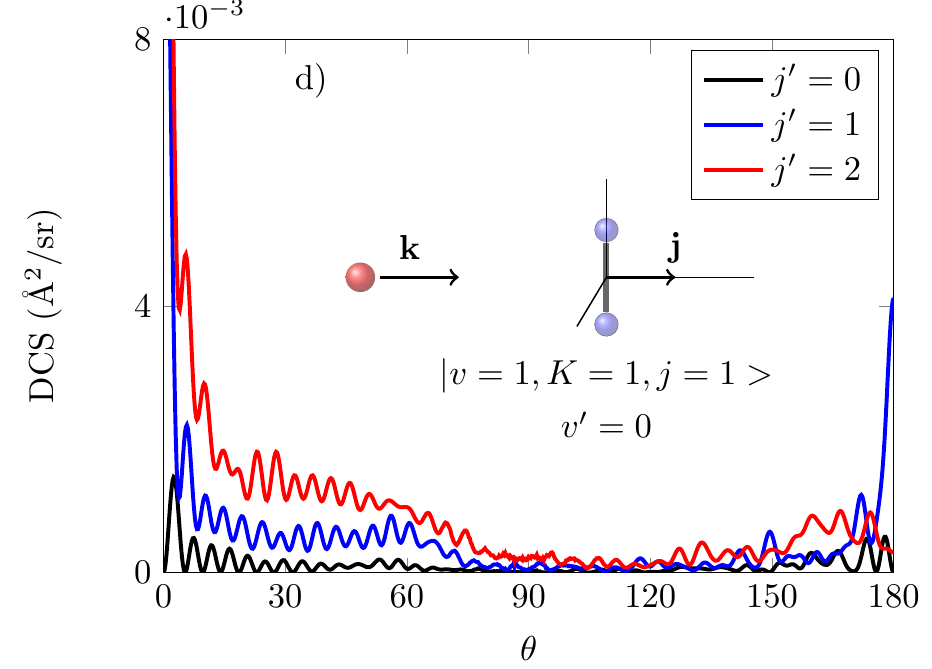} % Here is how to import EPS art
\caption{ \label{fig:stereody} Stereodynamics calculations of the DCSs for the proton transfer reaction between $\rm{H^{+}_2(v,j=1)}$ 
and He. Here we compute the contribution of the initial helicity quantum numbers $K=0$ and $K=1$ to the 
total state-to-state DCSs. The total collision energy is 1.5 eV in all cases. }
\end{figure}

The mechanism of the reaction can also be understood by finding the preferred direction 
of attack between the reactants. 
We have already mentioned in Section \ref{sec:pes} that the geometrical configuration between $\rm{He}$ and $\rm{H}_2^{+}$ during initial 
approach seems to be a key aspect in the reactive process. 
In an earlier study of ours on the inverse reaction \cite{Bovino:12}, it was suggested 
that differently rotationally excited reagents could selectively drive the 
products either through an ``abstraction'' mechanism or an ``insertion'' mechanism, 
with different angular distributions of the corresponding products.

The effects caused on the present reaction by the initial orientation between the fragments  
can therefore also be studied by performing stereodynamics calculations. 
In Figure \ref{fig:stereody} we present the contribution to the total DCSs of two different 
rotational states of o-$\rm{H}_{2}^{+}(j=1)$ in collisions occurring at 1.07 
(\ref{fig:stereody}.a and \ref{fig:stereody}.b) and 0.822 (\ref{fig:stereody}.c and \ref{fig:stereody}.d) eV of collision energies. 
Figure (\ref{fig:stereody}.a) shows the approach between $\rm{He}$ and  o-$\rm{H}_2^{+}$ in the rotational state described 
by the  helicity quantum number $K=0$. In this situation,  the ion o-$\rm{H}_2^{+}$ has 
the tendency to be orientated in the same direction of the initial relative velocity vector $\mathbf{k}$. 
The reaction is favored in this case by the minimum of the PES in the linear geometrical configuration.  
Moreover, at the present high collision energy, the efficiency of the reaction is also enhanced by 
an effective transfer of linear momentum between the three atoms: after the collision with He, 
one hydrogen in turn hits  back its identical partner, thereby losing most of its  
extra momentum in the direction of attack. Thus, it becomes more likely to be trapped by the He partner. 
The second Figure \ref{fig:stereody}.b shows the contribution of the rotational 
states described by helicity quantum number $K= 1$.
In this scenario, the molecule has the propensity to be orientated perpendicularly to 
the direction of $\mathbf{k}$, and therefore 
the helium atom will face in its approach the leading repulsive wall of the T-shape configuration (see Figure \ref{fig:pes}).  This approach is certainly less efficient to promote the reaction and therefore their related reactive DCSs 
are one order of magnitude lower than those describing the abstraction mechanism in the panel \ref{fig:stereody}.a.  
Similar ideas may be employed to describe the reactions that occur when the reacting molecular ions are 
in their first excited vibrational state (Figure \ref{fig:stereody}.c and \ref{fig:stereody}.d ). 
In this case we also observe the reduction of the backward flux, an effect on the reactive scattering that 
was already described before in this section.

\section{Conclusions}

In this work we have simulated,   using ab initio quantum scattering calculations, 
experimental VM images of the proton transfer reaction between $\rm{He}$ and $\rm{H_{2}^{+}}$.
We have found that it is feasible to distinguish among experimental images of reactions associated with different  
initial rotational (or vibrational) quantum states of the reactants. 
The experiments might then be able to verify some of the reaction mechanisms predicted by the quantum reactive calculations. The above findings could become experimentally possible as a result of the peculiar angular dependence of the reactive DCSs on the quantum initial states of the reacting molecular ion.  
We have observed, in fact, that the images of the VM broaden in the forward direction as the initial 
rotational quantum number of $\rm{H_{2}^{+}}$ increases, in agreement with the behaviour of the DCSs. 

We have further provided in our present discussion a series of qualitative interpretations of this 
behaviour by linking it to specific features of either the stereodynamics or the length of the times on interaction between reacting partners. From those considerations we were able to observe in the VM simulations that  the reactive flux increases into the backward scattering with the decreasing of the initial vibrational quantum numbers of $\rm{H_{2}^{+}}$.  Finally, we have additionally determined the geometrical configurations of maximal reactivity by performing stereodynamical quantum calculations.

In the future we intend to extend this study to the isotopologue $\rm{HD}^+$ in order to assess 
possibly detectable  isotope effects in the reaction with He. In this case, the combination of  VM experiments and  accurate calculations of the reactive DCSs  may reveal new interesting reaction mechanisms, thus  providing additional observables like the branching ratio of the reaction between the two available isotopic products.

\section{ACKNOWLEDGEMENTS} 

F.A.G. and R.W. thank the Austrian Science Fund (FWF) for 
supporting the present research through Project No. P29558-N36. 
We are also grateful to Bj{\"{o}}rn Bastian and Tim Michaelsen for many illuminating discussions 
and help on the preparation of the Velocity Map simulations.

% \nocite{*}
% \bibliographystyle{acm}
\bibliography{references}% Produces the bibliography via BibTeX.

\end{document}